\documentclass[preprint,notoc]{JHEP3}
\usepackage{epsfig}

\def\eslt{\not\!\!{E_T}}

\def\to{\rightarrow}

\def\bi{\begin{itemize}}
 \def\ei{\end{itemize}}

\def\c1p{C1^\prime}

\def\tg{\tilde g}

\def\tq{\tilde q}

\def\tw{\widetilde W}
\def\tz{\widetilde Z}

\def\be{\begin{equation}}  
\def\ee{\end{equation}}  
\def\bea{\begin{eqnarray}}  
\def\eea{\end{eqnarray}}  

\def\sps1ap{SPS1a$^\prime$}
\title{Discovery potential for SUSY\\
 at a high luminosity upgrade of LHC14 
}

\author{Howard Baer$^{a}$, V.~Barger$^b$, Andre Lessa$^c$ and Xerxes Tata$^{d}$\\
$^a$Dept.\ of Physics and Astronomy, University of Oklahoma, Norman, OK
  73019, USA\\
$b$ Physics Dept. University of Wisconsin, Madison, WI 53706, USA\\
$^c$Instituto de F\'isica, Universidade de S\~ao Paulo, S\~ao Paulo - SP, Brazil\\
$^d$Dept. of Physics and Astronomy, University of Hawaii, Honolulu, HI 96822, US\\
E-mail: \email{baer@nhn.ou.edu}, \email{barger@pheno.wisc.edu},
\email{lessa@fma.if.usp.br}, \email{tata@phys.hawaii.edu}}

\preprint{{CETUP*-12/003} \ {UH-511-1196-12}}

\abstract{After completion of the LHC8 run in 2012, the plan is to
  upgrade the LHC for operation close to its design energy 
$\sqrt{s}=14$ TeV, with a goal of collecting 
hundreds of fb$^{-1}$ of integrated luminosity. The time is propitious
to begin thinking of what is gained by even further LHC upgrades. In
this report, we compute an LHC14 reach for SUSY in the mSUGRA/CMSSM
model with an anticipated high luminosity upgrade.  We find that LHC14
with 300 (3000) fb$^{-1}$ has a reach for SUSY via gluino/squark
searches of $m_{\tg}\sim 3.2$ TeV ($3.6$ TeV) for $m_{\tq}\sim m_{\tg}$,
and a reach of $m_{\tg}\sim 1.8$ TeV (2.3 TeV) for $m_{\tq}\gg m_{\tg}$.
In the case where $m_{\tq}\gg m_{\tg}$, then the LHC14 reach for
chargino-neutralino production with decay into the $Wh+\eslt$ final
state reaches to $m_{\tg}\sim 2.6$ TeV for 3000 fb$^{-1}$.}

\keywords{Supersymmetry Phenomenology, Supersymmetric
  Standard Model, Large Hadron Collider}

\begin{document}

\section{Introduction}
\label{sec:intro}

The LHC collider has delivered $\sim 5$ fb$^{-1}$ of integrated
luminosity at $\sqrt{s}=7$ TeV (LHC7), and so far over $6$ fb$^{-1}$ at
8 TeV (LHC8).  These runs have met with great success as evidenced by a
$5\sigma$ discovery of a Higgs-like particle with $m_h\sim 125$ GeV.  So
far, no direct sign of supersymmetry (SUSY) has emerged, leading to mass
limits in the mSUGRA/CMSSM model\cite{msugra} of $m_{\tg}\gtrsim 1.4$ TeV
for $m_{\tq}\simeq m_{\tg}$ and $m_{\tg}\gtrsim 0.85$ TeV for $m_{\tq}\gg
m_{\tg}$ based on analyses of just LHC7 data.  LHC expects to continue
running through the remainder of 2012 with a goal of collecting $\sim
20$ fb$^{-1}$ at 8 TeV. In 2013-2014, LHC is expected to be shut down
for an energy upgrade, with running set to resume around 2015 with
$\sqrt{s}$ close to the LHC design energy of 14 TeV.  The goal then is
to amass of order hundreds of fb$^{-1}$ of integrated luminosity at
LHC14.

Planning has already begun for further upgrades beyond LHC14 with a
design luminosity $\sim 100$ fb$^{-1}$/yr.  One option is a possible
energy upgrade, which would require design, construction and deployment
of a completely new set of magnets. A more economical (and perhaps
technologically viable) alternative may be a luminosity upgrade, with
the possible target of gathering $\sim 3000$ fb$^{-1}$ of integrated
luminosity. In this short note, we try to quantify the increased reach
of LHC for SUSY if the total integrated luminosity is increased from 300
to 3000 fb$^{-1}$.  While the increasing sparticle mass limits from LHC
seem to make the mSUGRA model increasingly implausible (in light of
fine-tuning considerations), nonetheless we continue to work in this
paradigm case mainly for historical reasons:  moreover, many physicists are
familiar with the $m_0\ vs.\ m_{1/2}$ plane of this model,
it is easy to compare with projections from previous
studies\cite{lhcreach,lhc14,lhc7,update} and many current analyses of
data\cite{atlas,cms} continue to be presented in this framework.
In particular, in Ref. \cite{update}, the projected reach of LHC7 for
5-30~fb$^{-1}$ was calculated. The increase in beam energy to 8 TeV
should lead to a modest increase of expected reach beyond these
results. Drawing upon these results, we estimate that the LHC8 with
20~fb$^{-1}$ will probe out to $m_{\tg}\sim 1.8$ TeV for $m_{\tq}\simeq
m_{\tg}$ and to $m_{\tg}\sim 1$ TeV for $m_{\tq}\gg m_{\tg}$. At LHC14
with 100~fb$^{-1}$,
the gluino reach extends to 3.0~TeV if $m_{\tq}\simeq m_{\tg}$.

The preceeding LHC reach results have been obtained by looking for
signatures arising from gluino and squark pair production reactions
followed by cascade decays\cite{cascade}, leading to multijet plus
missing $E_T$ ($\eslt$) signatures along with possibly one or more
isolated leptons.\footnote{Tagging of $b$-jets may potentially increase the
LHC14 reach from above projections by as much as 20\% in the so-called
HB/FP region of parameter space\cite{btagstudies}.} It has been pointed
out long ago and emphasized more recently that in models with gaugino
mass unification, as higher sparticle masses are probed, ultimately
chargino and neutralino pair production reactions will dominate over
gluino and squark pair production. In these models -- where $|\mu |$ is
assumed much greater than gaugino masses $M_1$ and $M_2$ and where
$m_{\tq}\gg m_{\tg}$ with $m_{\tg}\gtrsim 1$ TeV -- the gaugino
production process, $pp\to\tw_1\tz_2$, tends to be the dominant
sparticle production cross section at LHC. At such high masses, the
dominant decay modes tend to be $\tw_1\to W\tz_1$ and $\tz_2\to \tz_1 h$
(also with some non-negligible fraction of $\tz_2\to\tz_1 Z$
decays). This has led some groups to consider the LHC reach in the
$pp\to\tw_1\tz_2\to W(\to l \nu) + h(\to b\bar{b})+\eslt$
channel\cite{wh}.

In this note, we will consider both the gluino and squark cascade decay
signatures and the $Wh+\eslt$ channel. A major problem in assessing the
LHC reach for extremely high integrated luminosity projections is to
gain reasonable background estimates from Standard Model (SM)
processes. As higher sparticle masses are probed, harder jet and $\eslt$
and other cuts are designed to maximize the reach for signal against
background.  With hard enough cuts, the expected SM background rates may
drop into the tens of events level, requiring simulations with up to
billions of events to attain the needed statistical accuracy.  Such large Monte Carlo
samples are highly time and space intensive at present.  Thus, in the
case of very hard cuts, here we resort to fits of SM background
projections which we hope will be within factors of a few times the real
result. We show details of our BG and signal generation in
Sec. \ref{sec:cuts}, along with our fits which are needed for very hard
cuts. In Sec. \ref{sec:reach} we conclude and summarize our results.

\section{LHC14 reach for SUSY with 300 fb$^{-1}$ and 3000 fb$^{-1}$}
\label{sec:cuts}

For the simulation of the background events, we use AlpGen\cite{alpgen}
to compute the hard scattering events and Pythia\cite{pythia} for the
subsequent showering and hadronization.  The Standard Model background
simulation details follows closely to the discussion in
Ref.~\cite{lhc14}, so we do not reproduce them here.  The only
difference is the inclusion of the $W(\rightarrow l \nu)+h$,
$Z(\rightarrow l^{+}l^{-} )+h$ and $t\bar{t} + h$ processes, where we
take $m_{h} = 125$ GeV.  The signal events were generated using Isajet
7.82\cite{isajet}. We assume the mSUGRA (CMSSM) framework\cite{msugra} with
$\tan\beta = 10$, $\mu > 0$ and $A_0 = -2m_{0}$; such a large negative
$A_0$ value ensures that $m_h\sim 123-127$ GeV throughout most of mSUGRA
parameter space\cite{h125}.  All $2\to 2$ SUSY production processes are
included at leading order.  To simulate detector efficiencies and
smearing, we use the toy detector simulation described in
Ref.~\cite{lhc14}. We assume the same detector parameters (including
$b$-tag effciency) for the 300 fb$^{-1}$ and 3000 fb$^{-1}$ scenarios.
Jets and isolated lepton are defined as follows:

\bi
\item Jets are hadronic clusters with $|\eta| < 3.0$,
$R\equiv\sqrt{\Delta\eta^2+\Delta\phi^2}\leq0.4$ and $E_T(jet)>50$ GeV.
\item Electrons and muons are considered isolated if they have $|\eta| <
2.0$, $p_T(l)>10 $ GeV with visible activity within a cone of $\Delta
R<0.2$ about the lepton direction, $\Sigma E_T^{cells}<5$ GeV.  
\item  We identify hadronic clusters as 
$b$-jets if they contain a $B$ hadron with $E_T(B)>$ 15 GeV, $\eta(B)<$ 3 and
$\Delta R(B,jet)<$ 0.5. We assume a tagging efficiency of 60$\%$ and 
light quark and gluon jets can be mis-tagged
as a $b$-jet with a probability 1/150 for $E_{T} \leq$ 100 GeV,
1/50 for $E_{T} \geq$ 250 GeV, with a linear interpolation
for 100 GeV $\leq E_{T} \leq$ 250 GeV \cite{btag} in between.
\ei

In order to address the discovery potential for distinct signal topologies, we investigate four different
channels:
\bi
\item $0l$: $n(l)=0$, $n(j) \ge 3$, $\{E_T(j_1),E_T(j_2),E_T(j_3)\} >
$\{100 GeV, 100 GeV, 50 GeV\};
\item $1l$: $n(l)=1$, $n(j) \ge 2$, $\{E_T(j_1),E_T(j_2)\} > $\{100 GeV,
100 GeV\};
\item $2l$: $n(l)=2$, $n(j) \ge 2$, $\{E_T(j_1),E_T(j_2)\} > $\{300 GeV,
300 GeV\};
\item $Wh$: $n(l)=1$, $n(b)= n(j)= 2$, $\Delta \phi(b,b) < \pi/2$,
$M_{eff} > 350$ GeV, $m_{T} > 125$ GeV, 100 GeV $< m_{bb} < $ 130 GeV;
\ei 
where $n(l)$ is the number of isolated leptons (electrons and
muons), $n(j)$ is the number of jets (including $b$-jets), $E_T(j_i)$ is
the transverse energy of the $i$-th jet, $n(b)$ is the number of
$b$-tagged jets, $\Delta \phi(b,b)$ is the azimuthal angle separation between
two $b$-jets, $M_{eff}=\sum_{i} E_T(j_i) + \sum_{i} p_T(l_i) + \eslt$,
$m_T$ is the transverse mass and $m_{bb}$ the invariant mass of the
b-jet pair.  While the $0l$, $1l$ and $2l$ channels focus mostly on
signal topologies from gluino and squark production and cascade decay, the $Wh$
channel targets $\tw_1 \tz_2$ production, with $\tw_1 \to W+\tz_1$ and
$\tz_2 \to h + \tz_1$, as discussed in \cite{wh}.\footnote{In the present study
  we have included the important $Z(\to \nu\bar{\nu})t\bar{t}$ background, which was not included
  in Ref. \cite{wh}, but becomes relevant for hard $\eslt$ cuts.}
Although these channels do not necessarily give the maximum reach in all regions of parameter space,
they are inclusive enough to discuss the gain of a luminosity upgrade.

For each of the above channels we plot the SM background and signal
$\eslt$ distributions and verify if the signal is visible for $\eslt >
\eslt_{cut}$, where the value of $\eslt_{cut}$ is allowed to vary in the
interval $0.1-1.5$ TeV (in steps of 0.1 TeV).  We deem that the signal
is visible if there is a value of $\eslt_{cut}$ such that, for $\eslt >
\eslt_{cut}$, the signal satisfies: \be SG \ge max\left[\ 5\ {\rm
events},\ 0.2BG,\ 5\sqrt{BG}\right],
\label{disc} 
\ee where $SG$ ($BG$) is the number of signal (background) events for a
given integrated luminosity.

Since $\eslt_{cut}$ can be as large as 1.5 TeV, there are large Monte
Carlo (MC) statistical uncertainties for such hard cuts, due to the
limited number of events in our background MC samples.  To reduce these
uncertainties, we {\it extrapolate} the background to large $\eslt$.
Since we will eventually require $\eslt > \eslt_{cut}$, it is more
convenient to consider the {\it cumulative} $\eslt$ distribution,
defined by \be \sigma(S_{miss}) \equiv \int^{\infty}_{S_{miss}}
\frac{d\sigma }{d \eslt} d\eslt .  \ee Thus, the total cross-section for
$\eslt > \eslt_{cut}$ is simply $\sigma(S_{miss} = \eslt_{cut})$.
Furthermore, if $d\sigma/d\eslt$ falls exponentially, so does
$\sigma(S_{miss})$.  Therefore we extrapolate the $\sigma(S_{miss})$
distribution to large $S_{miss}$ values, assuming an exponential shape
at large $S_{miss}$. The extrapolation of $\sigma(S_{miss})$ instead of
$d\sigma/d\eslt$ reduces the MC uncertainties, since the former is a
cumulative function.  As an example we show the $S_{miss}$ distribution
before and after the extrapolation for the $0l$ and $Wh$ channels in
Fig.~\ref{fig:smiss}.  As we can see, this extrapolation procedure
allows us to consider hard $\eslt$ cuts, which are essential for
isolating the signal at high integrated luminosities.
\FIGURE[tbh]{
\includegraphics[width=13cm,clip]{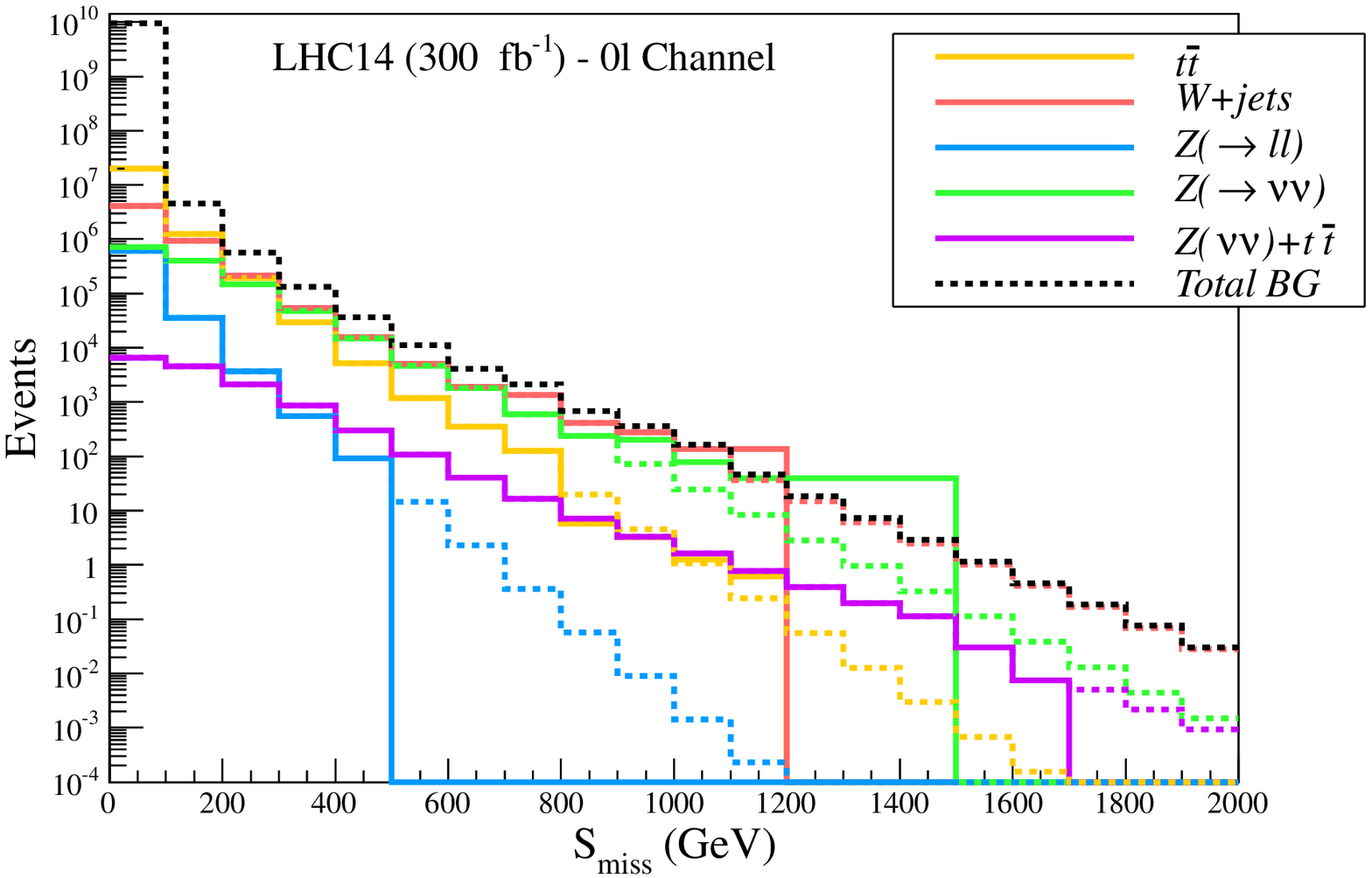}
\includegraphics[width=13cm,clip]{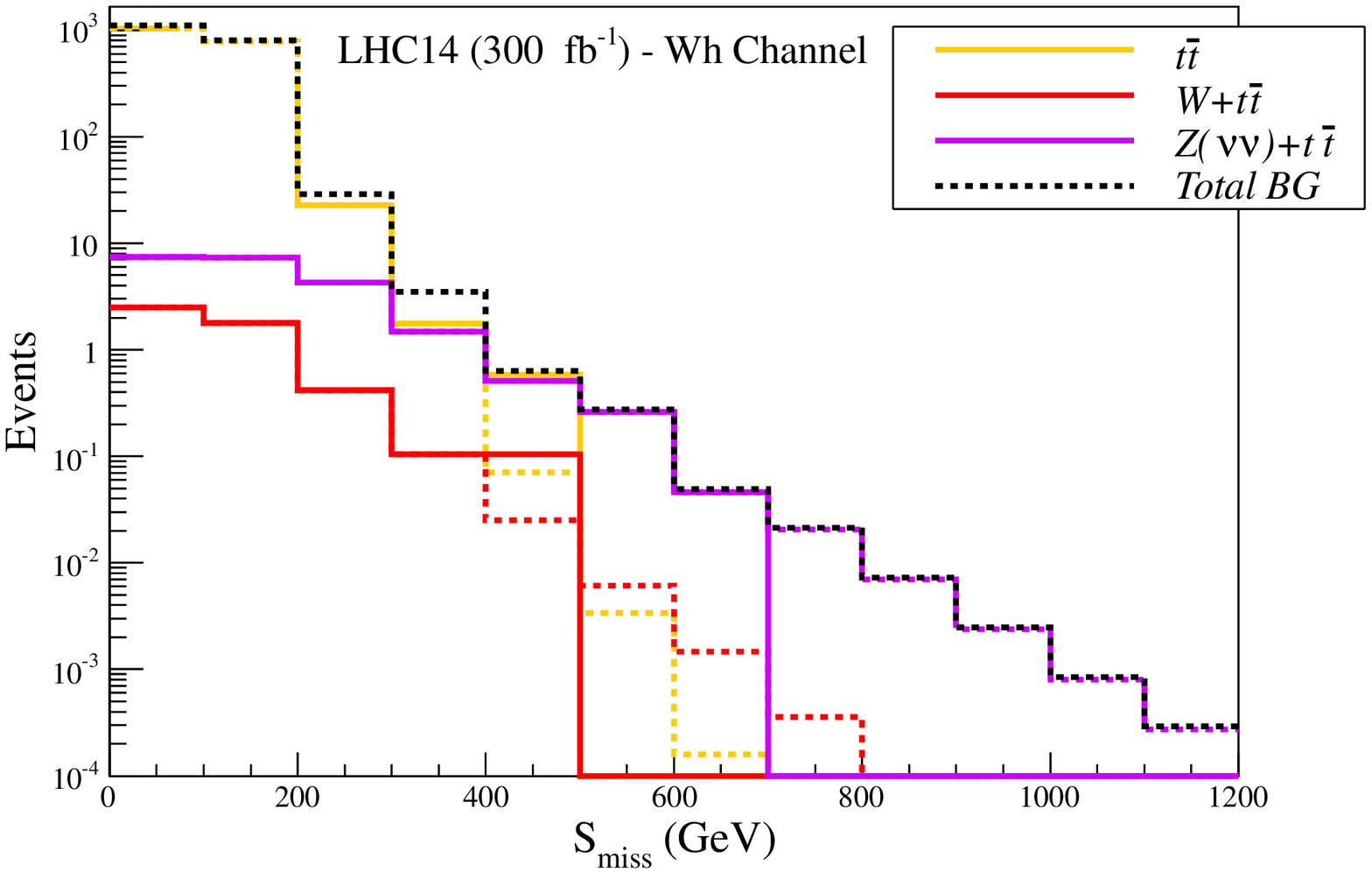}
\caption{Cumulative $\eslt$ distributions ($S_{miss}$) for the dominant
SM backgrounds in the $0l$ (top panel) and $Wh$ (bottom panel) channels
defined in the text.  We assume $\sqrt{s} = 14$ TeV and an integrated
luminosity of 300 fb$^{-1}$.  The solid lines represent the
distributions from our MC samples, while the dashed lines show the
extrapolated distribution used in our analysis, as discussed in the
text.  Only the dominant SM processes are shown.  The black dashed line
shows the total extrapolated background, which includes all SM
processes.}
\label{fig:smiss}}

Using the four channels listed above, as well as the extrapolated SM
background, we estimate the discovery potential for supersymmetry
assuming 300 fb$^{-1}$ and 3000 fb$^{-1}$ of integrated luminosity.  We
present our results in the $m_0\, vs.\, m_{1/2}$ plane and consider the 
$0l$, $1l$, $2l$ and $Wh$ channels separately.
We deliberately do not show results for the rate-limited but relatively
background-free same-sign dilepton and trilepton channels because we
were unable to reliably estimate the backgrounds for these high values
of integrated luminosity. Also, hard-to-estimate lepton fakes could make
substantial contributions to the background. 

Our results are shown in
Fig.~\ref{fig:reach}, where the solid lines show the reach in each
channel for 300 fb$^{-1}$, while the dashed lines correspond to the 3000
fb$^{-1}$ reach. The lower-left shaded region is excluded by SUSY searches at LHC7
using $\sim 5$ fb$^{-1}$ of data\cite{atlas,cms}.
\FIGURE[tbh]{
\includegraphics[width=13cm,clip]{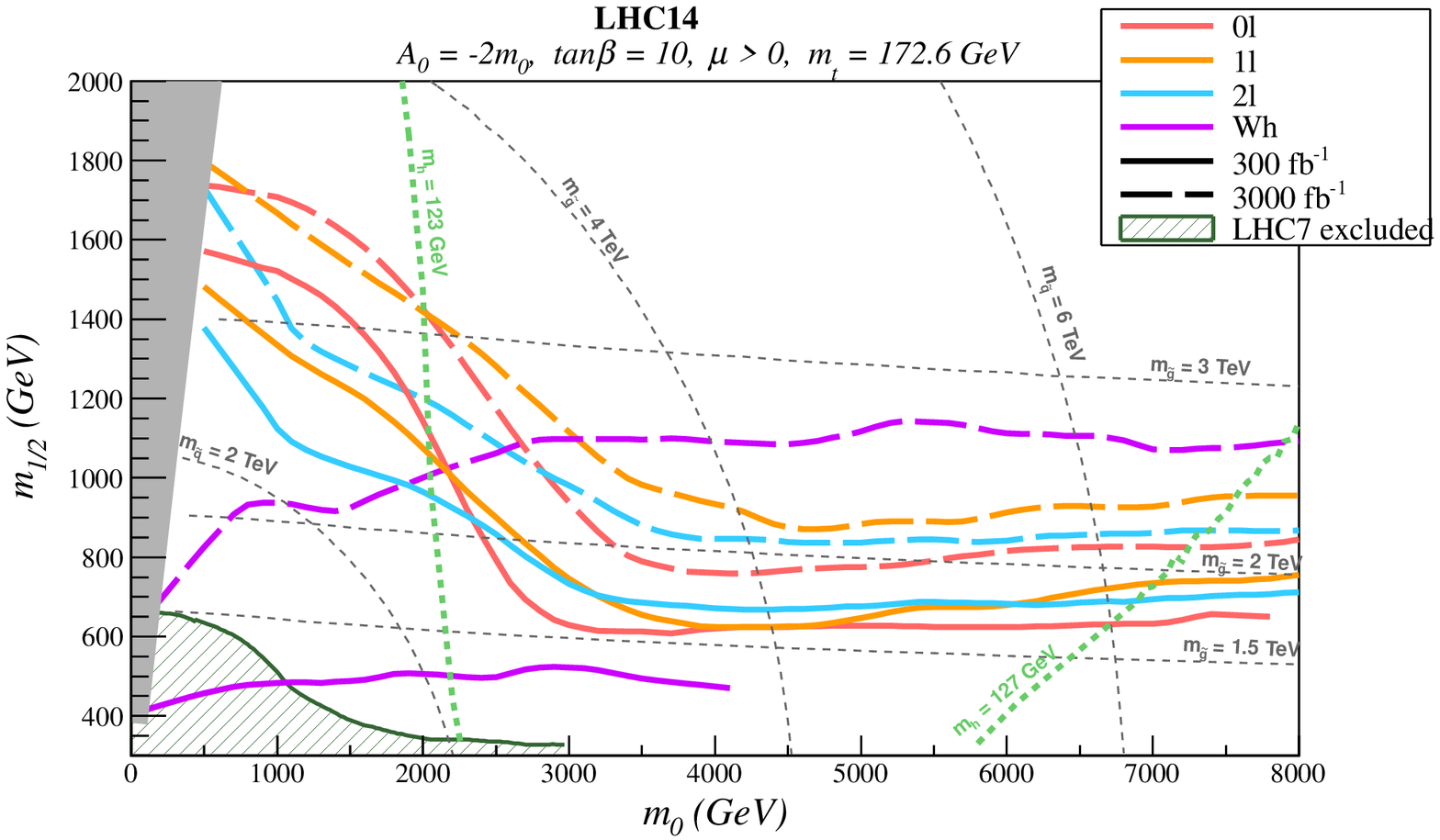}
\caption{SUSY reach in the four channels discussed in the text for LHC14
 for integrated luminosities of 300 fb$^{-1}$ (solid lines) and 3000
 fb$^{-1}$ (dashed lines).  The signal is observable if it falls below
 the curve for the corresponding integrated luminosity.  The fixed
 mSUGRA parameters are $A_0=-2m_0$, $\tan\beta =10$ and $\mu >0$.
 Gluino and squark mass contours are shown by the dashed, dark grey
 curves. We also show contours of $m_h=123$ and 127 GeV.  The shaded
 grey area on the left side of the figure is excluded because the stau
 becomes the LSP. The green shaded region in lower-left is excluded 
 by SUSY searches at LHC7.}
\label{fig:reach}}

For an integrated luminosity of 300 fb$^{-1}$, we see from
Fig.~\ref{fig:reach} that the $0l$ channel gives the maximum reach for
small $m_0$ values, where $m_{\tg} \sim m_{\tq}$.  In this case the
reach goes up to $m_{\tg} \sim 3.2$ TeV.  For higher $m_0$ values, where
$m_{\tg} \ll m_{\tq}$, the maximum reach is obtained in the $1l$ and
$2l$ channels and extends to $m_{\tg} \sim 1.8$ TeV.  At these mass
scales, the $\tw_1\tz_2\to Wh+\eslt$ channel gives a much smaller reach
at 300 fb$^{-1}$, going up to $m_{\tg} \sim$ 1.2 TeV in the squark
decoupling limit.

This picture significantly changes if we assume a high integrated luminosity of 3000 fb$^{-1}$. 
In this case, channels with smaller signal cross-sections, but larger signal/background ratios, such as the $1l$, $2l$ and $Wh$
channels, provide the maximum reach for most of the parameter space.
For $m_{\tq} \sim m_{\tg}$, the reach is dominated by the $0l$ and $1l$ channels and goes up to $m_{\tg} \sim $ 3.6 TeV.
For $m_0 \gtrsim$ 3 TeV, where squarks start to decouple, the maximum reach is obtained in the $Wh$ channel, since, for $m_{\tg} \gtrsim 2$ TeV,
electroweak gaugino production overcomes gluino production by almost an order of magnitude.

We see from Fig. 2 that for an integrated luminosity of 300 (3000)
fb$^{-1}$, the reach in the $Wh$ channel for $m_{\tg} \ll m_{\tq}$
extends to $m_{1/2} \sim 550$ GeV (1150 GeV) corresponding to $m_{\tw_1}
\sim 450$ GeV (950 GeV).  This reach corresponds to gluino masses of up
to $\sim 1.2$ (2.6) TeV.  Although these numbers superficially seem
lower than our earlier projections\cite{wh}, we should keep in mind that
here we have not included any signal K-factors and have also included
the $Zt\bar{t}$ background.  We note, however, that the
renormalization/factorization scale used here for the background
processes is such that the $t\bar{t}$ total cross-section is normalized
to its NLO value (for more details see Ref.\cite{lhc14}).

In Fig. \ref{fig:reach2}, we show the combined SUSY reach contours for
LHC14 with 100, 300, 1000 and 3000 fb$^{-1}$ of integrated luminosity.
The points below each curve are considered observable if they are
observable in at least one of the previously discussed four channels.
The kink in each of the 1 and 3~ab$^{-1}$ curves near $m_0\simeq 3-3.5$~TeV
occurs because the $Wh$ signal channel allows
one to probe larger $m_{1/2}$ than those accessible via gluino cascade
decays.

\FIGURE[tbh]{
\includegraphics[width=13cm,clip]{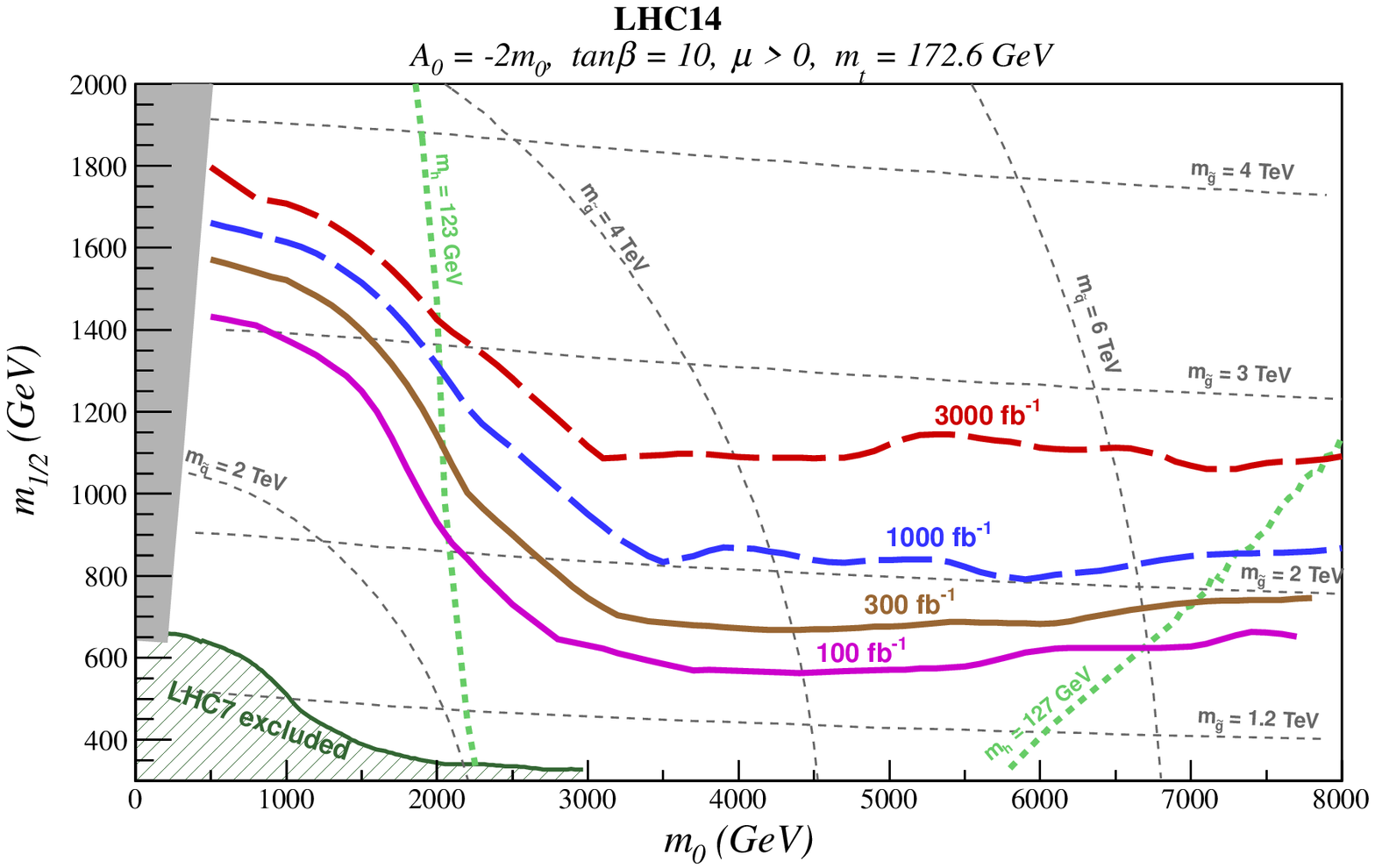}
\caption{SUSY reach in the {\it combined} four channels as 
 discussed in the text for LHC14
 for integrated luminosities of 100, 300, 1000 and 3000 fb$^{-1}$. 
 The signal is observable if it falls below
 the curve for the corresponding integrated luminosity. 
 The fixed  mSUGRA parameters are $A_0=-2m_0$, $\tan\beta =10$ and $\mu >0$.
 Gluino and squark mass contours are shown by the dashed, dark grey
 curves. We also show contours of $m_h=123$ and 127 GeV.  The shaded
 grey area on the left side of the figure is excluded because the stau
 becomes the LSP. The lower-left shaded region is excluded by SUSY searches at LHC7.}
\label{fig:reach2}}

\section{Conclusions}
\label{sec:reach}

In this study we have investigated the discovery potential of possible 
high luminosity upgrades of LHC14 for supersymmetry within the mSUGRA/CMSSM framework.
Previous reach projections for high integrated luminosity
values were presented in Ref.~\cite{lhc14}. In the current paper we
have made improved background projections/extrapolations involving the
cases where very hard cuts severely limit the statistical accuracy of
the background estimate. We have updated our mSUGRA projections with
$A_0\sim -2m_0$ so that the value of $m_h$ is close to 125 GeV
throughout most of parameter space.  Thirdly, we have included the reach
projection from $pp\to\tw_1\tz_2\to Wh+\eslt$ which should give the
dominant SUSY reach channel for $m_{\tq}\gg m_{\tg}$ and very high
integrated luminosity.

Our final reach projections listed in terms of $m_{\tg}$ in TeV units
are summarized in Table~\ref{tab:reach} for several integrated
luminosity values.  We find that LHC14 with 300 (3000) fb$^{-1}$ has a
reach for SUSY via gluino/squark searches of $m_{\tg}\sim 3.2$ TeV
($3.6$ TeV) for $m_{\tq}\sim m_{\tg}$, and a reach of $m_{\tg}\sim 1.8$
TeV (2.3 TeV) for $m_{\tq}\gg m_{\tg}$.  In the case where $m_{\tq}\gg
m_{\tg}$, the reach is higher in the $Wh$ channel, going up to $m_{\tg}
\sim 2.6$ TeV for 3000 fb$^{-1}$.  We point out that the reach in this
channel is only related to $m_{\tg}$ through the gaugino mass
unification assumption, since the $Wh$ channel depends only on the
$pp\to \tw_1 \tz_2$ production cross-section and the subsequent cascade
decays. For models where gaugino unification is not assumed, the reach is
independent of $m_{\tg}$ and goes up to $m_{\tw_1} \sim 450$ GeV (950
GeV), for 300 fb$^{-1}$ (3000 fb$^{-1}$) and $M_1 \leq M_2 \ll \mu$.

\begin{table}
\centering
\begin{tabular}{|c|c|c|c|}
\hline
IL (fb$^{-1}$) &  $m_{\tq}\sim m_{\tg}$ & $m_{\tq}\gg m_{\tg}$ & $Wh$   \\
\hline
100  &  3.0 TeV  & 1.6 TeV  & - TeV \\
300  &  3.2 TeV  & 1.8 TeV  & 1.2 TeV \\
1000 &  3.4 TeV  & 2.0 TeV  & 2.0 TeV \\
3000 &  3.6 TeV  & 2.3 TeV  & 2.6 TeV \\
\hline
\end{tabular}
\caption[]{Optimized SUSY reach of LHC14 within the mSUGRA model 
expressed in terms of $m_{\tg}$ for various choices of integrated luminosity.
The $m_{\tq}\sim m_{\tg}$ and $m_{\tq}\gg m_{\tg}$ values correspond to the maximum reach in the $0l$, $1l$ and $2l$ channels,
while the $Wh$ values correspond to the reach in the $Wh$ channel for $m_{\tq}\gg m_{\tg}$.
}
\label{tab:reach}
\end{table}

\section*{Acknowledgement} 
We thank A. Barr and A.Nisati for comments on the manuscript.
We thank the Center for Theoretical Underground Physics and Related Areas (CETUP* 2012) 
in South Dakota for its hospitality and for partial support during the completion 
of this work.
This research was supported in part by grants from the United States Department of Energy and Fundac\~ao de Apoio \`a Pesquisa do
Estado de S\~ao Paulo (FAPESP).

%

\end{document}